\documentclass[prl,floatfix,showpacs,superscriptaddress,longbibliography,twocolumn]{revtex4-2}

\usepackage[english]{babel}
\usepackage{amsmath,amssymb,amsfonts}
\usepackage{epsfig}
\usepackage{graphicx}
 \usepackage{relsize}
\usepackage{outlines}
\usepackage{tipa}
\usepackage{braket}
\usepackage[dvipsnames]{xcolor}

\usepackage[colorlinks=true,linkcolor=blue,citecolor=red]{hyperref}
\usepackage[normalem]{ulem}

\newcommand{\figu}[1]
{Fig.~\ref{#1}}

\begin{document}

\title{Spontaneous Formation of Exceptional Points at the Onset of Magnetism}
\author{L.~Crippa}
\affiliation{Institut f\"ur Theoretische Physik und Astrophysik and W\"urzburg-Dresden Cluster of Excellence ct.qmat, Universit\"at W\"urzburg, 97074 W\"urzburg, Germany}

\author{G.~Sangiovanni}
\affiliation{Institut f\"ur Theoretische Physik und Astrophysik and W\"urzburg-Dresden Cluster of Excellence ct.qmat, Universit\"at W\"urzburg, 97074 W\"urzburg, Germany}

\author{J.~C.~Budich}
\affiliation{Institute of Theoretical Physics, Technische Universit\"at Dresden and and W\"urzburg-Dresden Cluster of Excellence ct.qmat, 01062 Dresden, Germany}

\begin{abstract}
We reveal how symmetry-protected nodal points in topological semimetals may be promoted to pairs of generically stable exceptional points (EPs) by symmetry-breaking fluctuations at the onset of long-range order. This intriguing interplay between non-Hermitian (NH) topology and spontaneous symmetry breaking is exemplified by a magnetic NH Weyl phase spontaneously emerging at the surface of a strongly correlated three-dimensional topological insulator, when entering the ferromagnetic regime from a high-temperature paramagnetic phase. Here, electronic excitations with opposite spin acquire significantly different lifetimes, thus giving rise to an anti-Hermitian structure in spin that is incompatible with the chiral spin texture of the nodal surface states, and hence facilitates the spontaneous formation of EPs. We present numerical evidence of this phenomenon by solving a microscopic multi-band Hubbard model non-perturbatively in the framework of dynamical mean-field theory.    
\end{abstract}
\maketitle

In complex quantum many-body systems, physical properties are largely determined by correlation functions defining quasi-particles and their fundamental symmetries, rather than by elementary constituents such as bare electrons~\cite{Abrikosov1959}. The lifetime of quasi-particles, effectively modeled as an imaginary damping part of their energy, has played a crucial role in the literature for decades, e.g. for the understanding of Fermi- and Luttinger-liquids~\cite{Tomonaga1960,LuttingerWard1960,Luttinger1963}.
More recently, with the advent of non-Hermitian (NH) topological phases~\cite{ashida2020,Bergholtz2021}, numerous physical phenomena have been discovered that go conceptually beyond the simple damping of eigenmodes of an underlying Hermitian system~\cite{Rudner2009, Zeuner2015, Lee2016, Zhou2018, Lieu2018,  gong2018, Kunst2018, Yao2018, Yao2018b,nakagawa2018, Yuce2020, Xiao2020, Budich2020,Weidemann2020,Bergholtz2021,nobuyuki2021,yoshida2021}, prominently including exceptional points (EPs) as the generic NH counterpart of nodes in topological semimetals~\cite{kato1966,berry1994,Heiss2001,Heiss2012,Heiss2016,shen2018a,Yang2021,peters2021,staalhammar2021}. For such points to manifest, it is crucial that the NH self-energy $\Sigma$, describing the dissipation of a mode from its interaction with other degrees of freedom, acquires a matrix structure that is incompatible (non-commuting) with the free eigenmodes of the system~\cite{kozii2017,Bergholtz2021,rausch2021}. To this end, various systems have been considered that explicitly break a symmetry between some degrees of freedom such as spin or orbital, e.g. by a matrix structure in their interactions that is inherited by $\Sigma$~\cite{yoshida2020b,EP42021}. 

\begin{figure}[ht]
  \includegraphics[width=\linewidth]{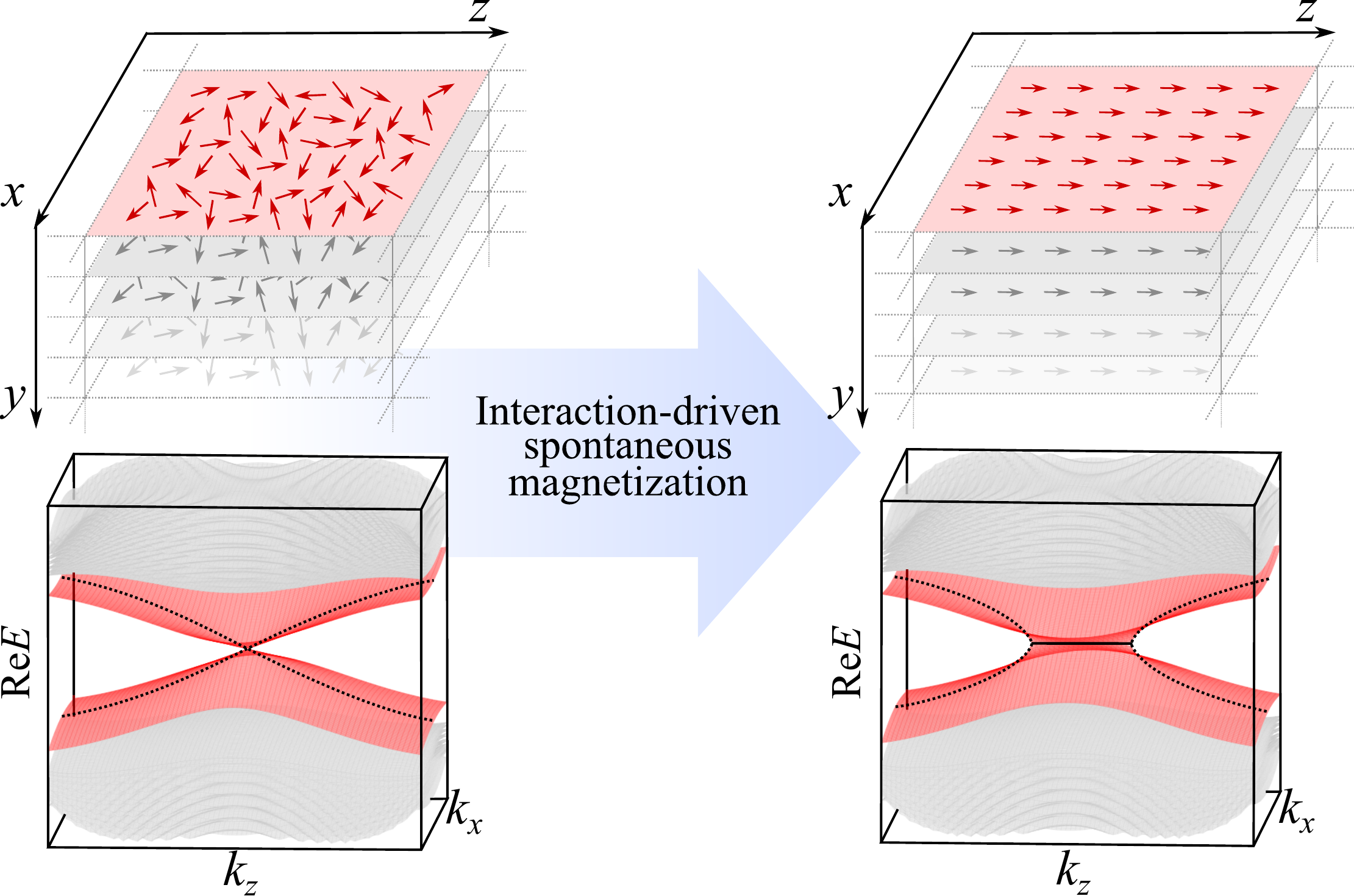}
    \caption{Formation of surface EPs from ferromagnetic fluctuations in a topological insulator setup with slab-geometry. In the noninteracting case, a paramagnetic surface semimetal with Dirac dispersion is found around high-symmetry points in the Brillouin Zone (left). Adding correlations, in a range of intermediate temperatures the onset of ferromagnetism with small $z$-magnetization and sizable magnetic fluctuations stabilizes a non-Hermitian Weyl phase (right). There, the quasi-particle dispersion exhibits a pair of exceptional points along the $k_{z}$ direction, from which a characteristic square root-dispersion emanates, and which are connected by a flat non-Hermitian Fermi arc.
  }
  \label{sketch}
\end{figure}

In this work, we reveal a profound interplay between spontaneous symmetry breaking and NH topology: close to an ordering transition, symmetry breaking fluctuations may \emph{spontaneously} give rise to a non-commuting anti-Hermitian structure, thus inducing topologically stable NH phenomena despite symmetric microscopic couplings.
As a case in point, we present numerical evidence that the symmetry-protected nodal surface states of a three-dimensional (3D) topological insulator (TI) can be promoted to a topologically robust magnetic NH Weyl phase featuring EPs (see Fig.~\ref{sketch} for an illustration). This occurs at intermediate temperature, where magnetic fluctuations induce a sizable anti-Hermitian spin dependence, in between a bulk ferromagnetic low-temperature phase (strong Hermitian spin dependence) and a paramagnetic high temperature phase (no spin dependence). While the ferromagnetic ground-state may be understood at Hartree-Fock level, the predicted NH Weyl phase with its exceptional degeneracies clearly eludes a mean-field description.

To reveal this phenomenon, we solve a multi-orbital Hubbard model for a 3D TI employing dynamical mean-field theory (DMFT)~\cite{georges1992,MullerHartmann1989,GeorgesKotliar1996}. In this framework, the ferromagnetic susceptibility at fractional filling as well as local dynamical fluctuations giving rise to anti-Hermitian self-energy contributions are well captured. We extract an effective quasi-particle Hamiltonian, compute and analyze the phase diagram of the system, and characterize the magnetic NH Weyl phase. As a spectroscopic hallmark, we find an increased spectral weight at the Fermi level, signaling an exceptional metallic dispersion rather than the Dirac semimetallic one of the noninteracting system.
We argue that our present findings exemplify a generic route towards NH topological phases, where the crucial ingredients are (i) symmetry-breaking fluctuations at the onset of long-range order, and (ii) the vicinity to a symmetry-protected nodal phase in the underlying free system. 

\noindent\textit{Exceptional points from magnetic fluctuations.} --
We consider a Hermitian quantum-many body system of electrons in which correlations give rise to non-Hermitian effects at the single-particle level, namely via the imaginary scattering rates entering the self-energy $\Sigma$. Such NH quasi-particle properties are conveniently encoded in the \textit{effective Hamiltonian}, defined as $H_{\mathrm{eff}}=H(\mathbf{k})+\Sigma(\mathbf{k},\omega)$~\cite{kozii2017,rausch2021}. We will be mainly interested in the physics around the Fermi level, i.e. at $\omega=0$ in our notation. Here, a NH contribution to the self-energy $\Sigma(\mathbf{k},0)$ arises at finite temperature as a finite lifetime of the single-particle excitations \cite{Abrikosov1959,kozii2017,Yoshida2018}. 

In the absence of spontaneous symmetry breaking, the self-energy will possess the symmetries that are not explicitly broken by the electron-electron interaction. In the symmetric regime both spin species have the same lifetime, thus excluding the formation of genuine NH phenomena such as EPs in spin-space. In contrast, in a long-range ordered phase such as a ferromagnet, a nontrivial $\Sigma$-matrix structure in spin may spontaneously emerge and, as we demonstrate in this work, stabilize EPs. We consider the case where, due to a spontaneous magnetization along the $\hat{z}$ direction, the self-energy matrix can be expressed in spin space as a linear combination $\delta\sigma_{0}+\gamma\sigma_{z}$, with $\delta$, $\gamma$ complex due to finite temperature and correlation strength.
While $\delta$ amounts to a constant (complex) energy shift, a finite $\gamma$ can generate exceptional points along the $k_{z}$ direction when added to a linearized Dirac Hamiltonian of the type $\mathbf{d}\cdot\overrightarrow{\sigma}$, where $\mathbf{d}\propto[k_z,0,k_x]$ \cite{Bergholtz2019,Bergholtz2021}. Generalizing from this minimal example, magnetic fluctuations with a strong in-plane component are capable of inducing exceptional points in 2D Dirac systems, as will be shown in the following within a microscopic lattice model.

\noindent\textit{Symmetry protected nodal Hamiltonian}-- One of the two crucial ingredients to the formation of EPs through ferromagnetism is a symmetry-protected 2D Dirac dispersion. To achieve it, we start from the 3D Bernevig-Hughes-Zhang (BHZ) model, described in reciprocal space by the Bloch Hamiltonian~\cite{Zhang2009,Liu2010}
\begin{equation}
\begin{split}
H_{\mathrm{BHZ}}(\mathbf{k})=&\mathcal{M}(\mathbf{k}) \sigma_{0}\otimes\tau_{z} +\lambda(\mathrm{sin}k_{x} \sigma_{z}\otimes\tau_{x}\\ & + \mathrm{sin}k_{y} \sigma_{0}\otimes\tau_{y} + \mathrm{sin}k_{z} \sigma_{x}\otimes\tau_{x})
\end{split}
\label{3dbhz}
\end{equation}
where $\mathcal{M}(\mathbf{k})=M-\epsilon(\cos{k_{x}}+\cos{k_{y}}+\cos{k_{z}})$ is the scalar expression for the diagonal hopping term and $M$ is a constant local splitting between two opposite-parity orbitals. The Pauli matrices $\sigma_{\{0,x,y,z\}}$ and $\tau_{\{0,x,y,z\}}$ act on the spin and orbital subspaces, respectively. We choose $\epsilon=1$ as the unit of energy, and set the spin-orbit coupling strength to $\lambda = 0.3\epsilon$.
In the following, we consider the case $M=0$, where the two orbitals are equally occupied: differently from the 2D BHZ model, the system is still insulating for this parameter set, and realizes a Weak Topological Insulator (WTI) phase~\cite{FuKaneMele2007} protected by time-reversal symmetry.
We study a finite slab geometry of the model, finite along the the spatial $y$-direction: in this setup, the Hamiltonian of the BHZ slab~\cite{Zhang2009} will show gapless surface states crossing at the high-symmetry points $X=[\pi,0]$ and $Z=[0,\pi]$. In the proximity of such points, the dispersion is linear in $k_{x}$ and $k_{z}$, and coupled to the Dirac matrices $\sigma_{z}\otimes\tau_{x}$ and  $\sigma_{x}\otimes\tau_{x}$ respectively. In particular, in the spin subspace it is of the characteristic Dirac form $\mathbf{d}\cdot\overrightarrow{\sigma}$, where $\mathbf{d}=[\pm k_z,0,\mp k_x]$. We note that the surface states of the WTI are neither protected against TRS-breaking perturbations, nor against weak disorder which entails a cell-doubling and projects the surface Dirac points on top of each other~\cite{FuKane2007PRB}.

From the form of the bulk Hamiltonian, it is easy to see that a NH perturbation of the type $i\zeta\sigma_{z}\otimes\tau_{x}$ (for real $\zeta$) in spin-orbital space guarantees the presence of exceptional points. Explicitly, the Hamiltonian in the spin subspace becomes non-diagonalizable if and only if $k_{x}=0$ and $k_{z}=\pm \zeta$, thereby realizing two $\mathrm{EPs}$ around the $X$ and $Z$ high symmetry points along the $k_{z}$ direction. 
We can further simplify the form of such a NH perturbation term in a slab geometry that is thick enough that the exponentially decaying surface states do not hybridize significantly~\cite{Zhang2009,Liu2010}. Under these conditions, the projected Hamiltonian of the surface states is of the general form $H_{\mathrm{surf}}=\lambda\langle \psi_{0} | \tau_{x} | \psi_{0} \rangle (\sigma_{x}k_{z}+\sigma_{z}k_{x})$, where $\psi_{0}$ is the 2-component spin up/down part of the TRS-related 4-component surface eigenstates $\Psi_{\uparrow}=[\psi_{0},0]$ and $\Psi_{\downarrow}=[0,\psi_{0}]$~\cite{Note1}.
The orbital subspace structure is now ``embedded" in a finite scalar coefficient, and a simple imaginary ferromagnetic term of the form $\sigma_{z}\otimes\tau_{0}$ is equally effective in generating EPs. 
\begin{figure}[h]
  \includegraphics[width=\linewidth]{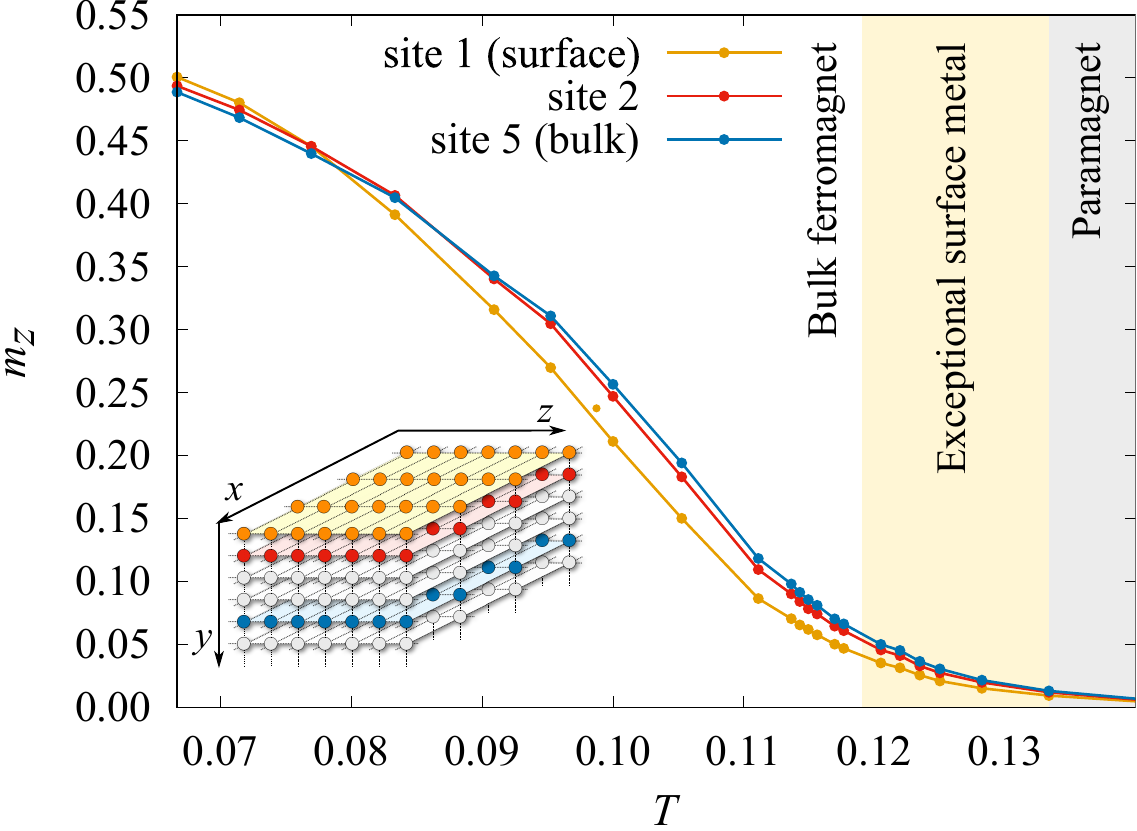}
    \caption{(color online) Magnetization $m_z$ as a function of $T$ for fixed filling $n=0.39, U=13, U'=7, J=6$.  The yellow shaded region represents the area where the magnetization is low enough to allow for EPs to appear at the Fermi energy. At larger $T$ (grey shaded), the system is in a paramagnetic phase. At smaller $T$ (white region), the system is ferromagnetic, but the difference in the real part of the spin-dependent self-energies is so high as to surmount the topological bandgap.
  }
  \label{mags}
\end{figure}

\noindent\textit{Interaction-mediated ferromagnetism}--
The second ingredient necessary to stabilize EPs is here represented by magnetic fluctuations.
In strongly correlated electron systems, it is well established that Hund's coupling can stabilize ferromagnetic phases in a wide range of fractional fillings for multi-orbital models~\cite{Vollhardt1999,Held1998}. 
However, the above BHZ slab model does not belong to this class of systems, as the surface Dirac point is pinned at half-filling. There, for large enough $U$, an antiferromagnetic phase is expected to develop instead~\cite{Kubo1975,Yoshida2013,amaricci2016}. To identify candidate systems that are susceptible to ferromagnetic order, we note that numerous compounds in both 2D and 3D exist that exhibit Dirac or Weyl physics away from half-filling, e.g. in the kagome and pyrochlore lattice family~\cite{Guo2009,Berke2018,Lim2020,Li2021}. 
As a minimal model for driving the Dirac points away from half-filling, we amend the 3D BHZ Hamiltonian of Eq.~(\ref{3dbhz}) by adding a third orbital with a trivial cosine dispersion. We refer to this additional orbital with the label $1$, while the two BHZ orbitals are $2$ and $3$, respectively. In this simple realization, orbital $1$ is coupled to the others only through the many-body interaction term $H_\mathrm{int}$.
The full Bloch Hamiltonian for the system then becomes
\begin{equation}
H(\mathbf{k})= \mathcal{M}_{1}(\mathbf{k}) \sigma_{0} \oplus H_{BHZ}(\mathbf{k}),
\label{3bands}
\end{equation}
where $\mathcal{M}_{1}(\mathbf{k})$ is defined analogously to $\mathcal M(\mathbf{k})$ (see Eq.\eqref{3dbhz}) with parameters $M_{1}=-3.0\epsilon$ and  $\epsilon_{1}=0.1\epsilon$, so as to be far from the Fermi level. To account for electronic interactions, we choose the full Hubbard-Kanamori potential~\cite{Kanamori1963,georges2013}, which in real space can be written as

\begin{equation}
\begin{split}
&H_{\mathrm{int}}=U\sum_{il}n_{il\uparrow}n_{il\downarrow}+\sum_{i,l<l^{'},\sigma\sigma^{'}}(U'-J\delta_{\sigma\sigma^{'}})n_{il\sigma}n_{il^{'}\sigma^{'}}\\
&-J\sum_{il\neq l^{'}}(c^{\dagger}_{il\uparrow}{c}_{il\downarrow}c^{\dagger}_{il^{'}\downarrow}c_{il^{'}\uparrow})+J\sum_{il\neq l^{'}}(c^{\dagger}_{il\uparrow}{c}^{\dagger}_{il\downarrow}c_{il^{'}\downarrow}c_{il^{'}\uparrow})
\end{split}
\label{HubbardKanamori}
\end{equation}
where $i$ represents the site and $l,l^{'}$ the orbitals. The first two terms refer to the density-density electron-electron interaction and Hund's coupling, while the last two terms describe spin-flip and pair-hopping processes.

This represents the most general two-body interaction form for a 3-orbital setup. In particular, we choose the Hund's coupling $J$, of crucial importance for the insurgence of ferromagnetism, to be rather large. The specific set of interaction parameters used in the following is thus $(U,U',J) = (13.0, 7.0, 6.0)$ in units of $\epsilon$.

This strongly correlated model is treated numerically making use of DMFT in its real-space formulation, solved by continuous time quantum Monte Carlo (CTQMC)~\cite{GeorgesKotliar1996,Gull2011,Potthoff1999,Schueler2017,Wallerberger2019}. Here, every layer of the system in the finite spatial direction is mapped to an inequivalent impurity model coupled to a bath. This approximation captures all local quantum fluctuations of the interacting lattice problem and the resulting self-energy is purely local. Notwithstanding the absence of momentum dependent and inter-layer correlations beyond mean field, real-space DMFT is expected to be well justified for our analysis: as recently shown with numerical and analytical cluster methods on similar models~\cite{localNonlocal2021}, nonlocal correlation effects become relevant only when the contribution from the local self-energy is very small. Such regime is however far from the case of our ferromagnetic setup for which we can conclude that non-local self-energy components amount at most to modest perturbations, against which the predicted EPs are globally robust.
The slab we consider has 20 layers stacked along $\hat{y}$: at this thickness, the surface states are sufficiently decoupled from each other and the self-energy of the internal layers settles to a uniform bulk value, consistent with a ``thick slab'' geometry. This we have also confirmed by further increasing the number of layers results, finding only minimal quantitative changes.

To stabilize the ferromagnetic phase, the value of the chemical potential $\mu$ is adjusted to obtain an occupation between 0.3 and 0.4, i.e. sufficiently away from half filling~\cite{Held1998}. The temperature $T$ is of crucial importance both for the magnetic and the topological phase: on one hand, too high a value of temperature will destroy long-range order and yield a paramagnetic state, where $\mathrm{Im}\Sigma$ becomes spin-independent. On the other hand, if $T$ is too low the real (Hermitian) magnetization will be so large as to completely close the topological bandgap at the chemical potential, thus rendering the system a bulk ferromagnetic metal. The NH Weyl phase featuring EPs, then, resides in an intermediate temperature regime, where sizable magnetic fluctuations coexist with a moderate magnetization at the onset of ferromagnetism.

In ~\figu{mags} we show the site-resolved magnetization curve for our three-band model (see Eqs.~(\ref{3bands}-\ref{HubbardKanamori})) for fixed filling $n=0.39$. For the range of temperatures highlighted by the yellow area, coinciding with the smearing of the phase transition due to finite-size effects~\cite{Binder1987}, exceptional points exist that are located precisely at, or very near to the Fermi energy. For higher values of temperature T, the system is paramagnetic, while for lower T it is a ferromagnetic bulk metal. It is interesting to note the magnetization profile reversal happening between surface and bulk around $T \approx 0.085$: for higher temperatures, the magnetization of the surface decreases more rapidly than that of the bulk approaching the critical point. This effect, related to the spin fluctuations in each layer of the slab at high temperature, is in agreement both with both experiments~\cite{Liebermann1969,Walker1984} and theoretical simulations~\cite{wai2009}, but is completely unrelated to the EP physics. Indeed, it is sufficient for the magnetization to be finite and small enough throughout the system to stabilize them. 

\begin{figure}[ht]
  \includegraphics[width=\linewidth]{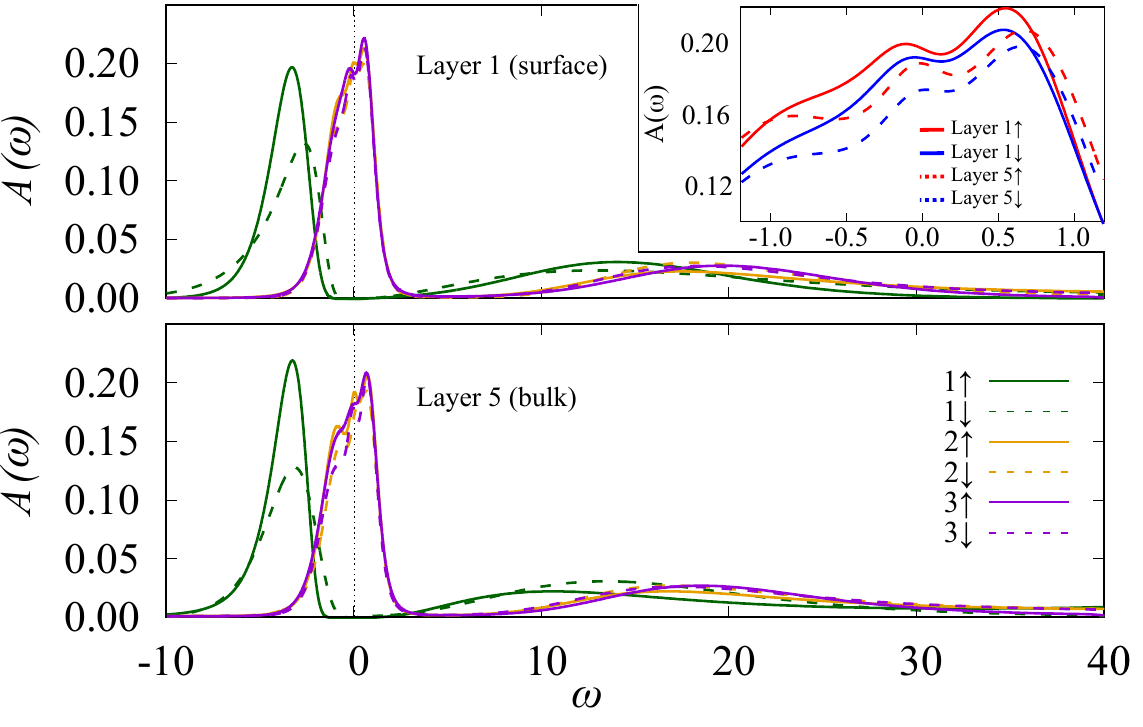}
    \caption{(color online) Site-resolved spectral function on the real frequency axis, for different orbitals (color) and spin (full/dashed line). The parameters of the model are $(U,U',J,T,\mu)=(13.0,7.0,6.0,0.13,2.00)$. 
    Top and bottom panels refer to the slab edge and bulk respectively.
    The auxiliary orbital $1$ (green) does not contribute spectral weight at the Fermi level, i.e. the physics is dominated by the EP and Fermi arc in the protected surface-state region. In the top inset, an averaged spectral weight for orbitals 2 and 3, resolved for spin and site, is plotted around the Fermi level, comparing the surface layer (solid lines) and a bulk layer (dashed lines).
  }
  \label{aw}
\end{figure}

\noindent\textit{Hallmarks of the EPs}-- We now analyze a series of properties that make the exceptional character of the surface metallic phase manifest. A first glimpse of the  EP physics comes by analyzing the site-resolved spectral function $A(\omega)$ on the real frequency axis, shown in \figu{aw}. The overall occupation of the model is around $n\approx0.39$, well within the stable ferromagnetic phase.

The data are obtained from the DMFT Matsubara local Green's function through maximum entropy analytical continuation~\cite{maxent2021}. Comparing the density profiles of orbitals $2$ and $3$ at the surface (top panel) and in the bulk (bottom panel) the presence of additional spectral weight in the first case is clearly visible. At the Fermi level, $A(\omega)$ features a finite difference between surface and bulk, as shown by the inset. This suggests that the Fermi surface of the interacting system has measure grater than 0, i.e. it doesn't consist simply of a discrete set of points, as is the case for the noninteracting semimetal.
Furthermore, \figu{aw} shows that the weight distribution of the auxiliary orbital $1$ is divided into two distinct regions: a broad peak below the Fermi level, in correspondence with the non-interacting band, and an incoherent dome at higher energy. Orbital 1 does not therefore contribute to the spectral weight at the Fermi level, and the physics of the exceptional points fully characterizes the behavior of the system at the given filling.

\begin{figure}[ht]
  \includegraphics[width=\linewidth]{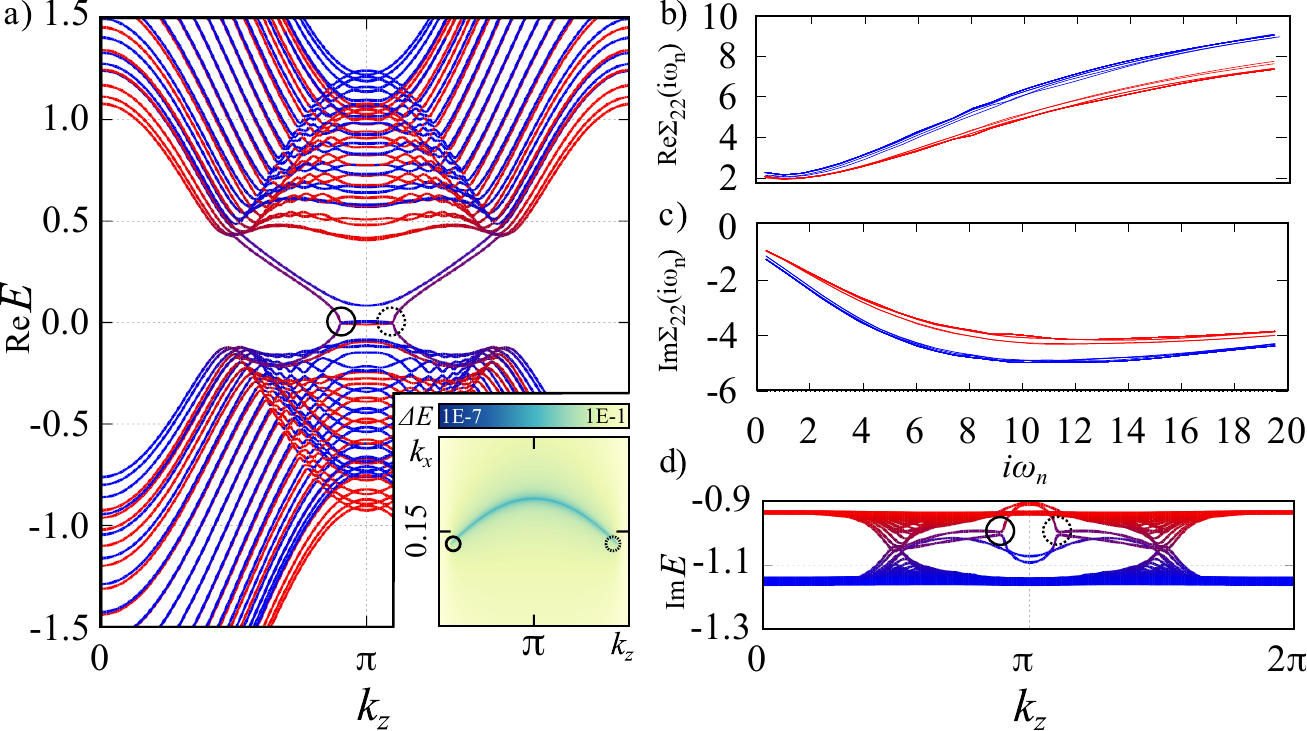}
    \caption{(color online) Effective Hamiltonian and self-energies for parameters $(U,U',J,T,\mu)=(13.0,7.0,6.0,0.13,2.00)$ \textbf{a}) Real part of the eigenvalue dispersion of the effective Hamiltonian near the Fermi level. The bands are plotted along the $k_{z}$ direction for the $k_{x}$ value at which the EPs sit. The two EPs have positive (solid circle) and negative (dotted circle) vorticity $\nu$, respectively. Between the EPs, the real energy dispersion of the surface bands is zero along a line (non-Hermitian Fermi arc) whose shape is shown in the inset.
    \textbf{b-c}) Real and imaginary parts of the orbital-diagonal self-energy of orbital $2$ on the Matsubara axis (orbital $3$ is analogous). Color encodes spin, and the multiple lines refer to the different planes of the slab. 
    \textbf{d}) Imaginary part of the eigenvalue dispersion of the effective Hamiltonian. Between the EPs, the energy dispersion of the surface bands is purely imaginary.
  }
  \label{bands}
\end{figure}

The most evident confirmation of the presence of EPs comes by studying the complex eigenvalue dispersion of the effective Hamiltonian at zero frequency $H_{\mathrm{eff}}(\mathbf{k})=H(\mathbf{k})+\Sigma(\mathbf{k},\omega=0)$~\cite{WangZhang2012,kozii2017,shen2018a}. \figu{bands}a shows the real energy dispersion of $H_\mathrm{eff}$ along a momentum direction parallel to $k_{z}$ in the two-dimensional Brillouin Zone. 
We only plot the bands possessing orbital character $2$ and $3$ in an energy window near the Fermi level, where the relevant physics is concentrated. As a consequence of the spin-dependent self-energy (a diagonal component of which is plotted in \figu{bands}b and c as an example), distinctive NH topological features emerge in the effective bandstructure: the surface bands, which traverse the topological bandgap, acquire a characteristic square-root dispersion~\cite{Bergholtz2021,kozii2017}, and touch at the two exceptional points, highlighted by the black circles. In the momentum region between the EPs, a zero-energy dispersion is found: this is the NH Fermi arc, responsible for the increased spectral weight of the system at the Fermi level. Here, the surface state eigenvalues of $H_{\mathrm{eff}}$ are purely imaginary (see \figu{bands}d).
To confirm the topological protection of the EPs, we calculate the associated invariant known as the \textit{vorticity} $\nu$: this quantity, unique to the non-Hermitian realm, is defined in the space of complex quasi-particle energies rather than the manifold of eigenvectors, and takes the form of a winding number of the energy spectrum~\cite{shen2018a,yoshida2020b}. For our finite-size system, it can be readily obtained from the dispersion of the surface bands via the expression
\begin{equation}
\nu=\dfrac{1}{2\pi}\oint d^{2}\mathbf{k}\mathbf{\nabla}_{\mathbf{k}}\mathrm{arg}[\Delta E(\mathbf{k})]
\label{vorticity}
\end{equation}
where $\Delta E(\mathbf{k})$ is the complex energy difference at the given $k$-point and the integral is done along a path encircling the EP.
By numerically performing the integration, we can confirm the topological protection of the left and right EP-candidates in \figu{bands}, which possess $\pm 1/2$ vorticity respectively. 
We can further assess the robustness of the EPs via a final observation: while a self-energy proportional to $i\sigma_{z}$ splits the Hermitian Dirac cone into two EPs that are connected by a Fermi arc along $k_{z}$, the real and orbital off-diagonal parts of $\Sigma$ generally do not favor this mechanism. However, such terms only manifest in a shift of the EPs along the $k_{x}$ direction~\footnote{See Supplemental Material at [URL will be inserted by publisher] for further details on the surface state Hamiltonian, effective band dispersion and full characterization of the topological gap.} and in a curvature of the Fermi arcs connecting them (see the inset of \figu{bands}a), highlighting the topological robustness of EPs.

\noindent\textit{Conclusion}-- We have demonstrated how symmetry-protected surface Dirac cones can be promoted to pairs of generically stable exceptional points by means of symmetry-breaking fluctuations. As a case in point, we have discussed an extended weak topological insulator model, the nodal surface states of which split into EPs as a consequence of ferromagnetic fluctuations. Ferromagnetic ordering has been facilitated by augmenting the model by a third band so as to drive the Dirac points away from half filling. The topological protection of the EPs strengthens the gapless surface phase of the system, which in the non-interacting limit is susceptible to both time-reversal symmetry breaking and weak disorder. Furthermore, the free semi-metallic spectral density at the surface is enhanced to a metallic one due to the presence of Fermi arcs connecting the EPs.

Conceptually different from previous predictions of EPs, our setup does not require any ad-hoc assumption on the orbital- or spin-dependence of the interaction parameters, nor does it build on coupling the system to environmental degrees of freedom. Instead, the nontrivial anti-Hermitian structure inducing the topologically stable exceptional points inherently emerges from spontaneous symmetry breaking.

Generalizations of our case study to other ordering transitions are readily conceivable. When spontaneous symmetry breaking selects the quantum number of some degree of freedom, a nontrivial matrix structure in this degree of freedom appears in correlation functions with respect to the symmetry broken state. There, fluctuations will generically induce a sizable anti-Hermitian contribution close to the transition that may stabilize EPs.

\noindent\textit{Acknowledgments} --  This work is funded by the Deutsche Forschungsgemeinschaft (DFG, German Research Foundation) through Project-ID 258499086-SFB 1170, Project-ID 247310070-SFB 1143 (Subproject A06), the W\"urzburg-Dresden Cluster of Excellence on Complexity and Topology in Quantum Matter –ct.qmat Project-ID 390858490-EXC 2147, and RE1469/13-1. The authors gratefully acknowledge the Gauss Centre for Supercomputing e.V. (www.gauss-centre.eu) for funding this project by providing computing time on the GCS Supercomputer SuperMUC at Leibniz Supercomputing Centre (www.lrz.de).

\bibliography{references}
\end{document}